\documentclass[12pt]{iopart}
\usepackage{graphicx}
\usepackage{tipa}
\usepackage{amssymb}
\begin{document}

\title[Mapping the frontiers of the nuclear mass surface]{Mapping the frontiers of the nuclear mass surface}

\author{Zach Meisel}

\address{Institute of Nuclear \& Particle Physics, Department of Physics \& Astronomy, Ohio University, Athens, OH, 45701 USA}
\ead{meisel@ohio.edu}
\vspace{10pt}
\begin{indented}
\item[]October 2019
\end{indented}

\begin{abstract}
Nuclear masses play a central role in nuclear astrophysics, significantly impacting
the origin of the elements and observables used to constrain ultradense matter. A
variety of techniques are available to meet this need, varying in their emphasis on
precision and reach from stability. Here I briefly summarize the status of and near-future for the time-of-flight magnetic-rigidity (TOF-$B\rho$) mass measurement technique, emphasizing the complementary and interconnectedness with higher-precision mass measurement methods. This includes of recent 
examples from TOF-$B\rho$ mass measurements that map the evolution of nuclear structure across the nuclear landscape and significantly impact the results and interpretation of astrophysical
model calculations. I also forecast expected expansion in the known nuclear mass surface from future measurement at the Facility for Rare Isotope Beams.
\end{abstract}

%
%
%
%
%

\section{Introduction}

Nuclear masses, and more specifically nuclear mass differences, are fundamental descriptors of atomic nuclei. Mass differences reflect the evolving energetics associated with changes in nuclear structure across the nuclear landscape as well as the energy costs (and gains) for nuclear reactions in astrophysical environments. Typical mass differences of interest are separation energies, e.g. for two-neutrons,
\begin{equation}
    S_{2n}(Z,N)=ME(Z,N-2)+2ME(0,1)-ME(Z,N)
    \label{eqn:Sn}
\end{equation}
and the reaction $Q$-value
\begin{equation}
    Q=\Sigma_{\rm reactants}ME - \Sigma_{\rm products}ME,
    \label{eqn:qvalue}
\end{equation}
where $ME$ is the atomic mass excess, $Z$ is the proton number, and $N$ is the neutron number. Changes in the slope of $S_{2n}$ for neutron-rich isotopes of an element provide signatures of neutron shell and subshell closures, while $Q$-values are essential inputs into astrophysics model calculations.

Therefore, nuclear mass measurements continue to play an essential role in nuclear physics studies, in particular for nuclear structure and nuclear astrophysics. Recent contributions include the emergence of the $N=32$~\cite{Leis18} and $N=34$ shell closures~\cite{Mich18}, mapping the island of inversion near $N=40$~\cite{Moug18}, $Q$-value determinations essential for calculations of type-I X-ray bursts~\cite{Ong18,Valv18}, and determining the trend in masses of neutron rich nuclei whose imprint can be seen in calculations of astrophysical $r$-process abundance patterns~\cite{Orfo18,Vill18}. 

\begin{figure} [ht]
\label{figure:crust}
\centering
\includegraphics[width=0.75\textwidth]{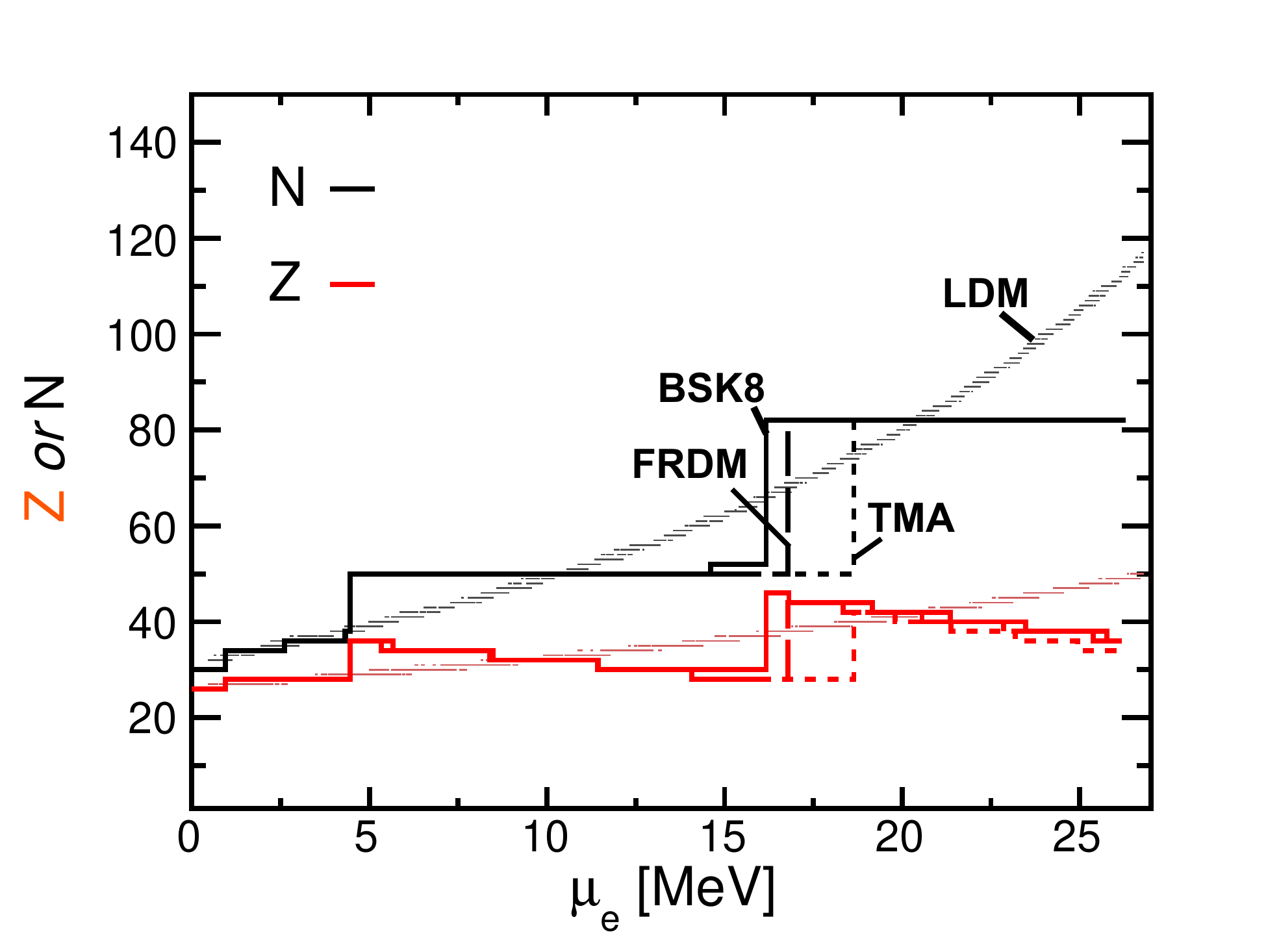}
\caption{Equilibrium neutron number $N$ and proton number $Z$ for depths in a non-accreting neutron star as indicated by the electron chemical potential $\mu_{e}$. Data are from References \cite{Meis18} and \cite{Rust06}.}
\end{figure}

In all of these cases, precise nuclear mass determinations were required to contribute to solving the problem at hand. However, it is important to note that ``precision" is a relative concept, where the necessary mass precision for a given scenario depends on the context. For instance, consider the case of $^{36}{\rm Ar}(p,\gamma)$, whose dependence on a single low-energy resonance means that keV-level changes to the $Q$-value leads to tens of percent changes in the astrophysical reaction rate~\cite{Ilia19}. Near the other extreme is the much lower precision required to constrain the properties of the neutron star crust. As an example, Figure~\ref{figure:crust} show the equilibrium composition for a cold-catalyzed neutron star crust using various nuclear mass models. Here MeV-level differences between nuclear masses are required to modify the onset of the $N=82$ shell closure, to which the composition converges deep in the outer crust.

The precision that can be achieved in nuclear mass measurement varies for the available measurement techniques, where the cost of increased precision is often increased measurement time. A map of achieved measurement precision for nuclides of various half-lives is shown in Figure~3 of Reference~\cite{Lunn19}. While a nuclear mass $m$ precision of $(\delta m)/m<10^{-8}$ has been achieved for penning trap mass spectrometry (PTMS), this is in general only possible for a half-life $t_{1/2}>1$~s. For the shortest $t_{1/2}$ and lowest production rates, the time-of-flight magnetic-rigidity (TOF-$B\rho$) method can be used, albeit at the cost of precision. The TOF-$B\rho$ method is generally limited to a precision of a few times $10^{-6}$. 

\begin{figure} [ht]
\centering
\includegraphics[width=0.9\textwidth]{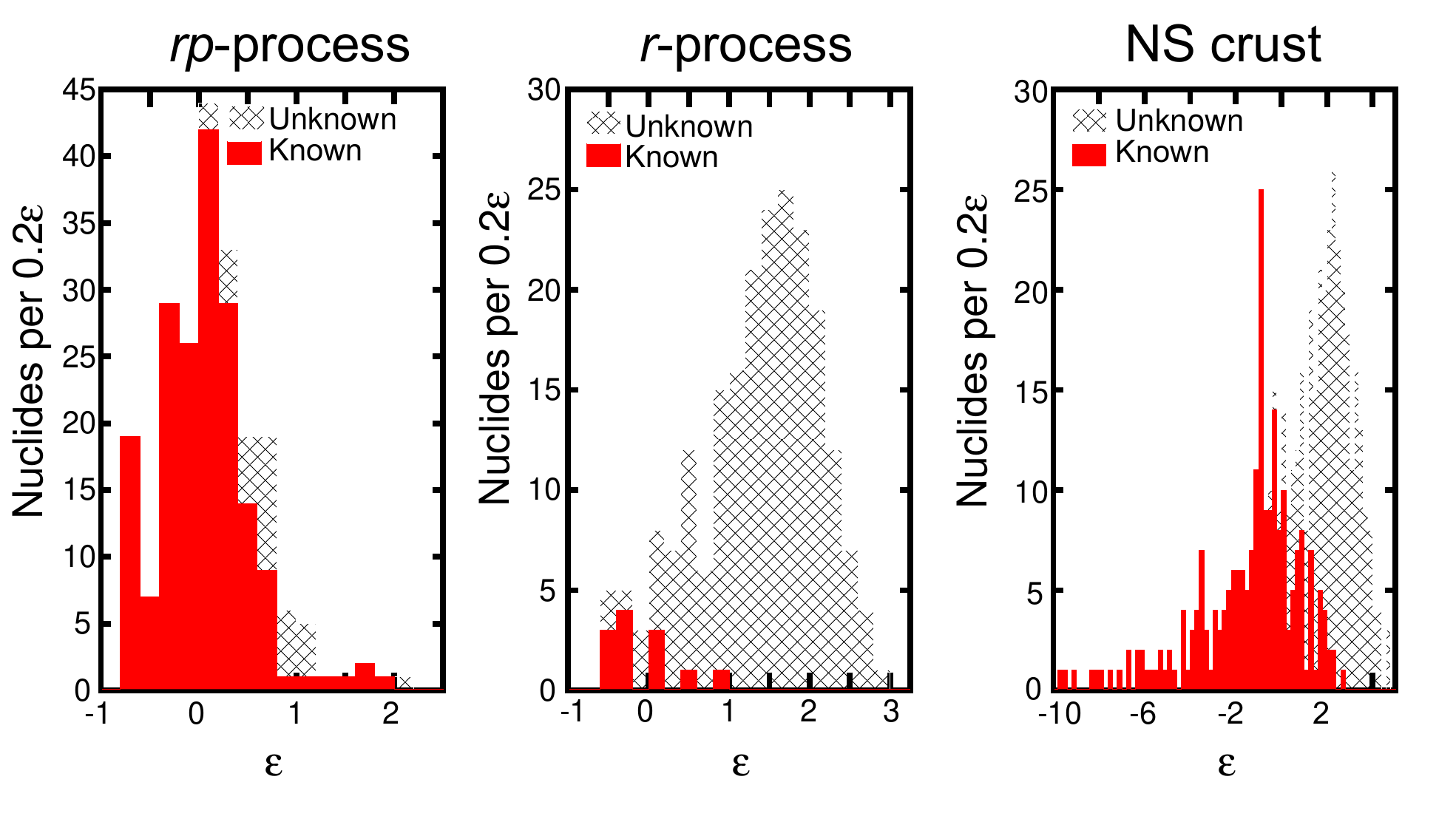}
\caption{Nuclides involved in astrophysical processes with known (solid-red) or unknown (hatched black) nuclear mass, binned by \textepsilon. The $rp$, $r$, and neutron star crust reaction network paths are from References \cite{Scha06}, \cite{Scha05}, and \cite{Lau18}, respectively.}
\label{figure:exoticity}
\end{figure}

The value of a relatively low-precision mass measurement technique can be seen by considering the characteristics of nuclides of interest for astrophysical processes. The difficulty of measuring a nuclide in the laboratory can be quantified using the exoticity~\cite{Meis13},

\begin{equation}
    \textrm{\textepsilon}=\log_{10}\left|\frac{dN_{\rm stab}}{t_{1/2}(dN_{\rm drip}+1)}\right|,
\end{equation}
where $dN_{\rm stab}$ is the number of neutrons from stability along an isotopic chain and $dN_{\rm drip}$ is the same for the neutron dripline, e.g. as defined by the FRDM~\cite{Moll95} mass model. Typically (see Figure 6 of Reference~\cite{Meis13}), PTMS is limited to \textepsilon$\,\,<1$, while TOF-$B\rho$ probes out to roughly \textepsilon$\,\,=4$. For context, consider the \textepsilon~distributions for nuclides involved in astrophysical processes shown in Figure~\ref{figure:exoticity}. It is clear that measurement techniques accessing \textepsilon$\,\,>1$ are essential to fully map the nuclear mass surface in the region of interest for astrophysical processes involving neutron-rich nuclides.

The remainder of this article focuses on the TOF-$B\rho$ mass measurement method. Section~\ref{sec:method} briefly summarizes the TOF-$B\rho$ method, Section~\ref{sec:contributions} highlights some significant contributions of TOF-$B\rho$ measurements to nuclear structure and nuclear astrophysics, and Section~\ref{sec:future} provides a preview of mass measurement achievements anticipated at the upcoming Facility for Rare Isotope Beams (FRIB).

\section{The TOF-$B\rho$ mass measurement method}
\label{sec:method}

The concept for the TOF-$B\rho$ method is that the nuclear mass can be determined by equating the centripetal and Lorentz forces on a charged massive particle (i.e. a nucleus) moving through a magnetic system. After some straightforward algebra and applying a relativistic correction in the form of the Lorentz factor $\gamma$, it is apparent that the rest mass
\begin{equation}
    m_{0}=\frac{{\rm TOF}}{L_{\rm path}}\frac{qB\rho}{\gamma},
    \label{eqn:tofbrho}
\end{equation}
where TOF is the time-of-flight along a path of length $L_{\rm path}$ for an ion with charge $q$ and magnetic rigidity $B\rho$. In practice, this relationship is not practicable for nuclear mass determinations of the necessary precision. For instance, for the typical conditions of a TOF-$B\rho$ experiment~\cite{Meis16}, using Equation \ref{eqn:tofbrho} to determine a nuclear mass to the $10^{-6}$-level would require knowing $L_{\rm path}\sim60$~m to tens of microns. Instead, an empirical relation is established by determining the $B\rho$-corrected TOF, TOF${\rm '}$, of several nuclides whose mass is known, e.g. from PTMS, to high-precision:
\begin{equation}
    m_{0}=f(Z,A=Z+N,{\rm TOF'}).
    \label{eqn:masseqn}
\end{equation}

Ultimately, the empirical approach requires determining the average TOF within less than a picosecond and $B\rho$ via a sub-millimeter measurement of the vertical displacement at a dispersive focus for several tens of nuclides. Ideally, the nuclides whose mass is known (``reference nuclides") have $Z$ and $A/Z$ similar to the nuclides of interest.

\begin{figure} [ht]
\centering
\includegraphics[width=0.45\textwidth]{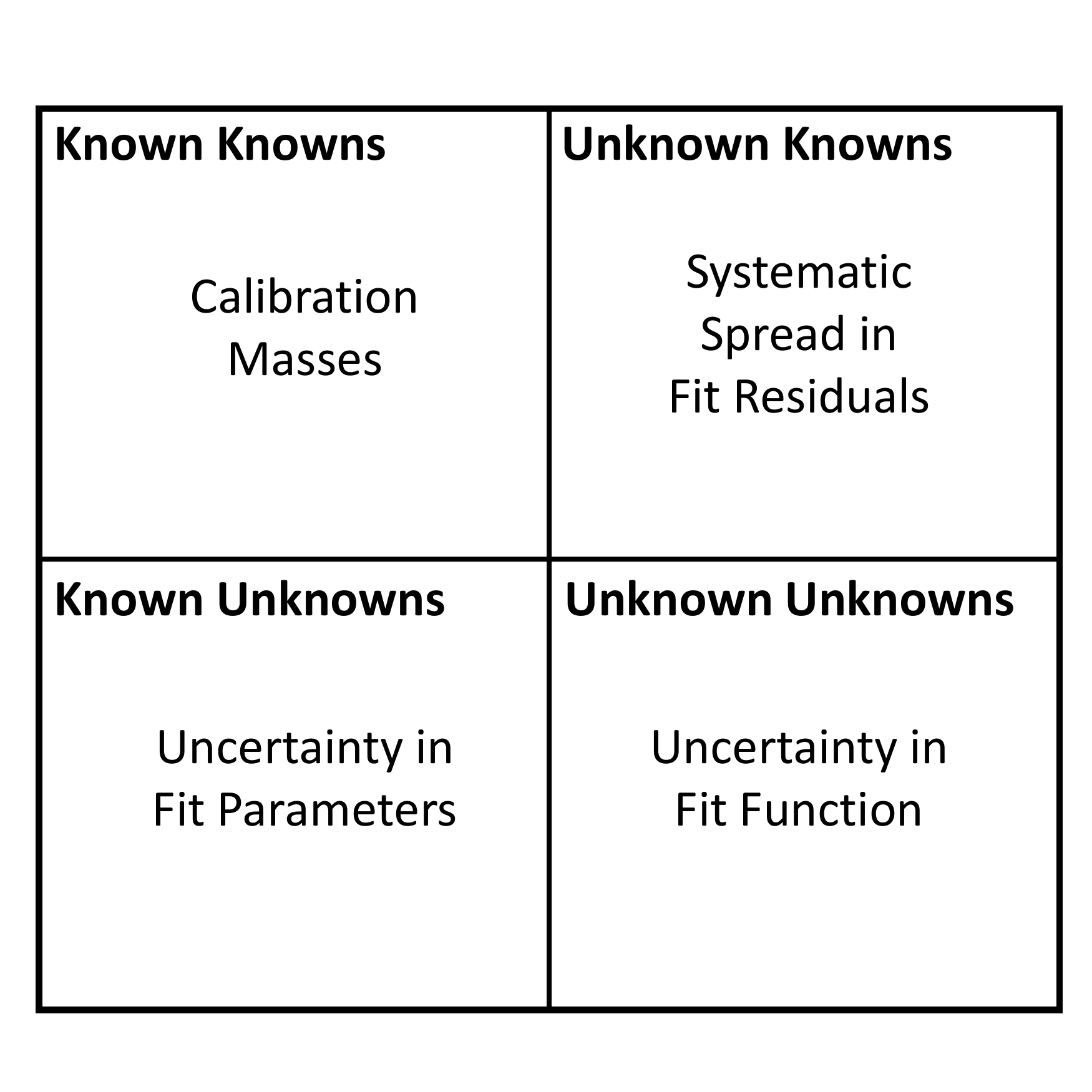}
\caption{TOF-$B\rho$ mass measurement uncertainty considerations categorized via a Rumsfeld Quadrant, following Reference~\cite{Stei12}.}
\label{figure:quadrant}
\end{figure}

Uncertainty quantification in the TOF-$B\rho$ method provides a significant challenge, due to the careful consideration required to assign systematic uncertainties. To aid in this discussion, consider the Rumsfeld Quadrant~\cite{Stei12} shown in Figure~\ref{figure:quadrant}.

Masses of reference nuclides provide known knowns through which to calibrate Equation~\ref{eqn:masseqn}. Naturally, the fit parameters for this equation have uncertainties which can be determined via standard error propagation, or, to ensure adequate accounting of numerous multi-collinearities, via a Monte Carlo technique as in Reference~\cite{Meis16}. These are the known unknowns. 

Systematic uncertainties of two distinct types often dominate the final mass uncertainty (for nuclides with greater than $\sim$500 counts). An unknown that we know of is the systematic spread remaining in the residuals of the fit to reference nuclides which nearly always have $\chi^{2}_{\nu}>1$. This uncertainty is generally accounted for by adding a blanket uncertainty in $m/q$ (as the actual fit function used determines this quantity) until $\chi^{2}_{\nu}=1$. The second, and unfortunately often omitted, systematic uncertainty comes from the unknown unknown: we do not know if the relation ultimately used in Equation~\ref{eqn:masseqn} is the fit-function that best describes the data. While Occam's razor dictates that the simplest model should be preferred, this simplicity needs to be balanced with the quality of the overall fit. A suggested approach is to use $\Delta\chi^{2}=\chi^{2}_{i}-\chi^{2}_{\rm min}$, where $\chi^{2}_{i}$ applies to a given model and $\chi^{2}_{\rm min}$ is the model resulting in the overall best-fit. The set of fit-functions which nominally describe the data equally well within some degree of confidence can be determined using $\Delta\chi^{2}$ tabulated for the number of degrees of freedom~\cite{Pres92}.

\section{TOF-$B\rho$ contributions to nuclear structure and nuclear astrophysics}
\label{sec:contributions}

\begin{figure} [ht]
\centering
\includegraphics[width=0.75\textwidth]{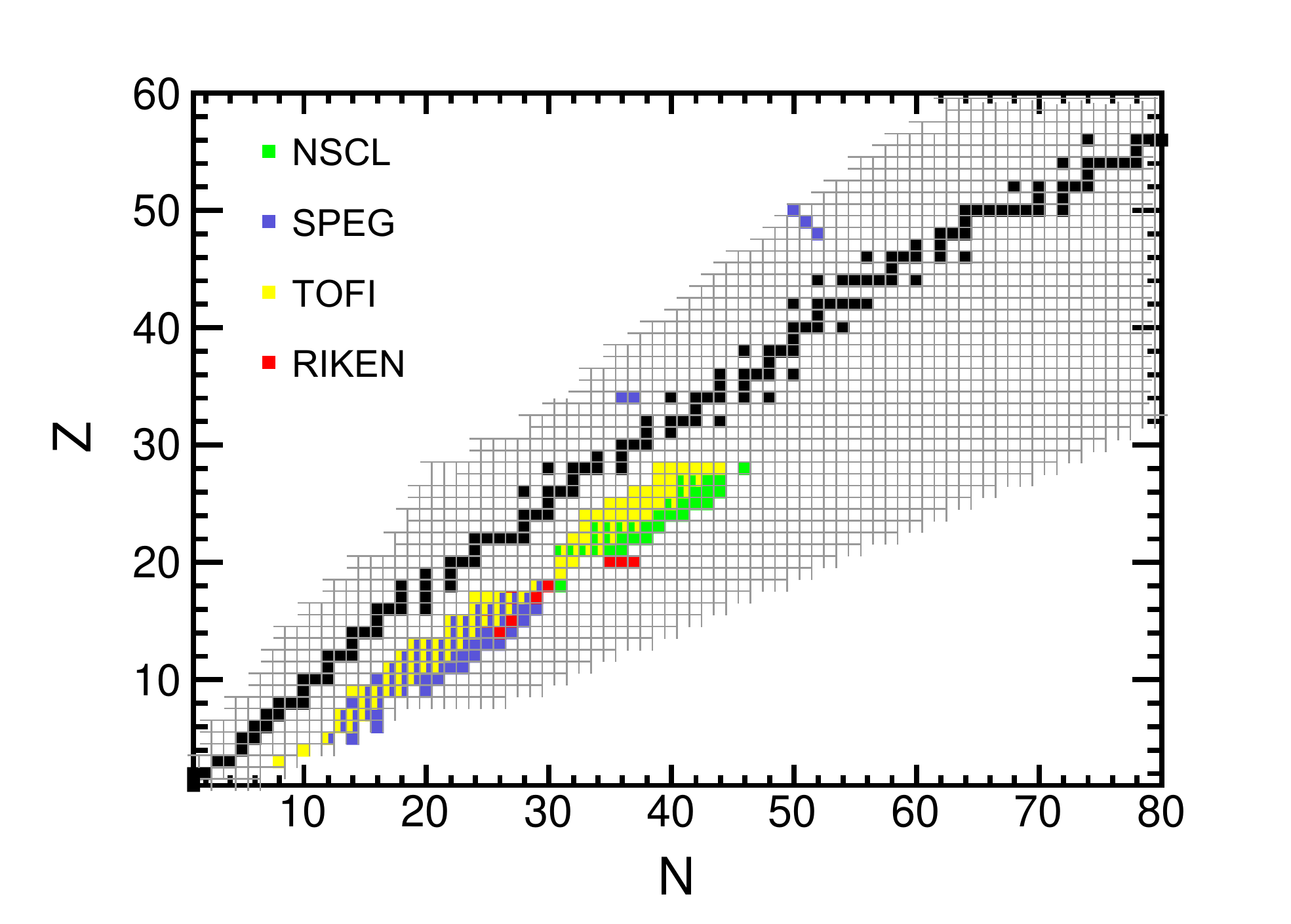}
\caption{Nuclear masses measured by the TOF-$B\rho$ method at the NSCL~\cite{Mato12}, SPEG at GANIL~\cite{Bian89}, TOFI at LAMPF~\cite{Wout85}, and RIKEN~\cite{Mich18}.}
\label{figure:facility}
\end{figure}

More than 300 nuclear masses have been determined using the TOF-$B\rho$ method, corresponding to a total of roughly 150 nuclides. Figure~\ref{figure:facility} shows these contributions by facility. Early measurements were performed using SPEG at GANIL~\cite{Bian89} and TOFI at the LAMPF~\cite{Wout85} facility, though neither set-up is still operational. In the past decade, TOF-$B\rho$ has been employed at NSCL~\cite{Mato12} and RIKEN~\cite{Mich18}. The concentration of measurements for $Z<30$ is largely due to the difficulty of dealing with multiple charge-states present for studies of higher-$Z$ nuclides, though efforts are ongoing to address this difficulty, e.g. using the technique of Reference~\cite{Tara09}.

Nuclear structure studies have identified regions of shape coexistence, the appearance and disappearance of shell closures, and the existence of halo nuclides. For instance, the lower-bound of the $N=28$ shell closure was mapped by References~\cite{Sara00,Sava05,Meis15}. The properties of halo nuclides were determined by References~\cite{Wout85,Gaud12}, where the latter was key to establishing the two-neutron halo nature of $^{22}{\rm C}$ and one-neutron halo of $^{31}{\rm Ne}$.

Achievements in nuclear astrophysics have largely focused on improving models of the accreted neutron star crust, where nuclear masses determine the location and strength of heat sources and heat sinks occurring due to electron-capture reactions~\cite{Meis18}. Thus far, measurement results~\cite{Meis16,Estr11} have indicated that electron-capture heat sources appear to be weaker than previously predicted. However, the accreted neutron star crust is not as cool as was once thought possible, since Reference~\cite{Meis15b} found that the strongest predicted heat sink in fact does not exist. Work is ongoing to expand TOF-$B\rho$ studies of astrophysical interest to the $r$-process region~\cite{Estr19}.

\section{The future of TOF-$B\rho$ mass measurements}
\label{sec:future}

While a handful of measurement targets remain at existing facilities, to significantly extend the TOF-$B\rho$ method to more exotic isotopes will require state-of-the-art radioactive ion beam facilities, such as FRIB. The purpose of this section is to forecast the extent to which the known nuclear mass surface is likely to be expanded at FRIB by coordinated efforts in TOF-$B\rho$ and PTMS measurements. This is done by discussing predicted FRIB production rates followed by anticipated achievements in PTMS, TOF-$B\rho$, and the two techniques combined.

\subsection{FRIB production rates}
\label{FRIBrates}

FRIB production rate predictions\footnote{At present,
calculations for individual nuclides can be obtained at
\tt{https://groups.nscl.msu.edu/frib/rates/fribrates.html}.} are calculated using
the software from
Reference~\cite{Boll11}, whose assumptions are briefly described
here. Fast-beam rates, required for TOF-B$\rho$, were calculated employing the KTUY
mass model~\cite{Kour05}, EPAX 2.15 fragmentation cross section
parameterization~\cite{Summ00},
LISE++3EER model for production cross sections from in-flight
fission~\cite{Tara05}, and LISE++v9.2.68 for beam transmission
efficiency~\cite{Tara08}. In each case the
rate chosen is that from the optimum primary beam, i.e. the one of
the 47 anticipated primary beams producing the highest rate, and a
beam power of 400~kW. For the case of stopped beams, which are
required for PTMS, the beam rate is reduced by the gas-stopping
efficiency ($\sim$1-50\%, depending on the ion mass and incident
beam-rate), the radioactive decay of ions during the 0.05~s
extraction time, and
the transport efficiency from the gas-stopper to the downstream
charge-breeder, which is assumed to be 80\%~\cite{Boll11}.

Fast beam rates of 
roughly $10^{8}$~particles per second (pps) are anticipated for nuclides
just beyond the present limit of known masses, where the production rate
generally drops off one order of magnitude for every 1-2 additional
neutrons from stability.
The stopped-beam rate is 10\% of the
fast-beam rate on average, but mostly ranges from 0-40\% (0\% cases are due
to short half-lives).

\subsection{Penning trap mass measurement}
\label{PTMS}

PTMS, the highest-precision mass
measurement technique presently available for rare isotopes, will be
performed with the Low Energy Beam Ion Trap (LEBIT) Penning trap at
FRIB~\cite{Ring13}. 
The PTMS technique consists of obtaining nuclear masses
by measuring the resonant frequency of the nucleus of
interest with respect to the resonant frequency for an ion (or
typically atomic cluster) of known mass orbiting within a
few cubic-centimeter volume, confined by a strong magnetic field and
hyperbolic electrodes~\cite{Blau13}. PTMS has been demonstrated to
deliver a mass measurement precision of $10^{-6}$
or better for as little as $\sim$50~measured ions~\cite{Mina12} and for nuclides
with half-lives as short as $\sim$10~ms~\cite{Smit08}.

The time-of-flight ion-cyclotron-resonance (TOF-ICR) technique is predominantly
employed for PTMS. For TOF-ICR PTMS, the cyclotron resonance of the ion in the trap,
which is directly proportional to its mass, is identified by
converting the orbital motion in the trap to a longitudinal motion
out of the trap and finding the minimum TOF to a fixed detection
location. The measurement uncertainty for TOF-ICR is reduced by
storing individual ions for long times in the trap and observing
several ions over a large enough frequency range to map the
cyclotron resonance. The relative statistical uncertainty $\delta
m/m$, which is generally
much larger than the systematic uncertainty for PTMS of rare
isotopes, is roughly given by  $\delta m/m\approx R^{-1}\rm{n}^{-1/2}$,
where n is the number of ions detected and $R$ is the resolving
power~\cite{Lunn03}. The resolving power is approximately equal to
the product of the cyclotron frequency of the ion in the trap
$f_{\rm{c}}$ (typically $\mathcal{O}\sim$~MHz) and the length of time the ion orbits in the trap
$t_{\rm{obs}}$ (typically $\mathcal{O}\sim$0.1~s). $R$ therefore
depends on many considerations, such as the mass of the nucleus of
interest, the obtainable charge-state, the time it takes to produce
the optimum charge state, and the nuclear half-life. 

Given the uncertainties in charge-breeding capabilities 
and the approximate nature of the estimate for n, I
make the approximation that $R=10^{5}$ for all nuclides of interest,
which is in-line with sample cases for rare isotopes~\cite{Boll01}.
For simplicity, I assume $t_{\rm{obs}}$=100~ms, and therefore
n will be the product of the stopped-beam rate and the duration of
the experiment, reduced due to the radioactive decay of ions during
the measurement process. For the experiment duration, I assume
24~hours of measurement time for the ion of interest, as is typical
for PTMS measurements of exotic nuclides. Experimental
$\beta$-decay half-lives from Reference~\cite{Tuli11} are used when available and
predictions from Reference~\cite{Moll03} are used otherwise. I
assume a systematic uncertainty typical for the measurement
precision of reference ions, $\delta m/m=10^{-8}$~\cite{Reds13}.

Above $A\approx50$, the known mass surface will be
extended by PTMS by a few isotopes or more for each isotopic chain.
The improvement over current nuclear mass uncertainties is also
significant, considering that many nuclear masses at the present
experimental frontier are only known to precisions of $\delta
m/m\approx10^{-5}$~\cite{Audi12}. The greatest gains are
expected for neutron-rich isotopes with $Z>55$, which will
substantially improve the predictive power of rare-earth element
nucleosynthesis in the $r$-process~\cite{Mump16}. Note that the
predicted gains on the proton-rich side are unreliable, as
particle-decays are not taken into account in our estimates.
Furthermore, the estimates neglect the potential existence of
isomeric states and isobaric contaminants~\cite{Kank12}; however,
recent improvements in PTMS, such as the stored
waveform inverse fourier transform~\cite{Kwia15b} and
phase-imaging ion-cyclotron-resonance~\cite{Elis15} 
techniques will mitigate the impact of these complications.
Additionally, the reach of PTMS may be extended by near-future
developments such as the single ion penning trap (SIPT)
method~\cite{Reds13}; however, expectations for SIPT have yet to be
benchmarked with rare isotope measurements and so are not considered
here.

\subsection{TOF-$B\rho$ mass measurement}
\label{TOFMS}

The TOF-$B\rho$ method is described in Section~\ref{sec:method}. The details of the uncertainty evaluation are elaborated upon here as they are pertinent to developing a forecast of anticipated measurement results.
It is important to note that the TOF-$B\rho$ method relies on the availability of PTMS results for
nuclides nearby (in terms of $Z$ and $N$) the isotopes of interest
and is primarily used as a tool to extend the known mass surface by
a few more neutrons along an isotopic chain.

Unlike PTMS, the measurement uncertainty of TOF-$B\rho$ is generally
dominated by systematic uncertainties due the many unknowns which
must be accounted for along the large experimental
set-ups~\cite{Meis16,Lunn03}. Therefore, the estimation technique for the
mass measurement uncertainty achievable via TOF-$B\rho$ is somewhat more
approximate. The statistical uncertainty of TOF-$B\rho$ is related to the
TOF measurement precision $\sigma_{\rm{TOF}}$/TOF and number of
measured ions n by $\delta
m/m\approx\sigma_{\rm{TOF}}/(\rm{TOF}\sqrt{n})$, where a typical
$\sigma_{\rm{TOF}}/\rm{TOF}$ of $10^{-4}$~\cite{Sava01,Estr11,Meis15} is assumed here.
I base the systematic uncertainty on the rough empirically
motivated
approximation\footnote{Note that the true systematic uncertainty
depends on many factors, not least the details of the local TOF-mass
relationship and the availability of suitable reference
nuclides~\cite{Meis16}.}~\cite{Meis16,Meis15,Gaud12,Estr11,Meis15b,Sava01}
that $\delta m/m|_{\rm{syst}}=5\times10^{-6}(1+(N-N_{\rm{ref}}))$,
where $N-N_{\rm{ref}}$ is the number of neutrons separating the
nuclide of interest and the most neutron-rich isotope of that
element with a mass uncertainty $\leq10^{-6}$. I sum the
statistical and systematic uncertainties to arrive at a total
uncertainty and assume a measurement time of 100~hours, as is
typical for recent TOF-$B\rho$ experiments.

In general TOF-$B\rho$ extends the measurable mass surface by 1-3 nuclides
with a precision useful to applications in nuclear structure and
nuclear astrophysics ($\lesssim$ a few times $10^{-5}$). The given
estimates ignore potential complications such as the existence of
isomers, multiple charge-states, and magnetic rigidity limits of
experimental equipment~\cite{Lunn03}; however, the former will
generally require further experimental work, techniques to deal with
multiple charge states~\cite{Tara09} have recently been implemented for TOF-$B\rho$~\cite{Estr19}, and it is anticipated that FRIB will host the high-rigidity spectrometer with a more than sufficient maximum rigidity, so I do not consider these complications further.

\begin{figure} [ht]
\centering
\includegraphics[width=1.0\textwidth]{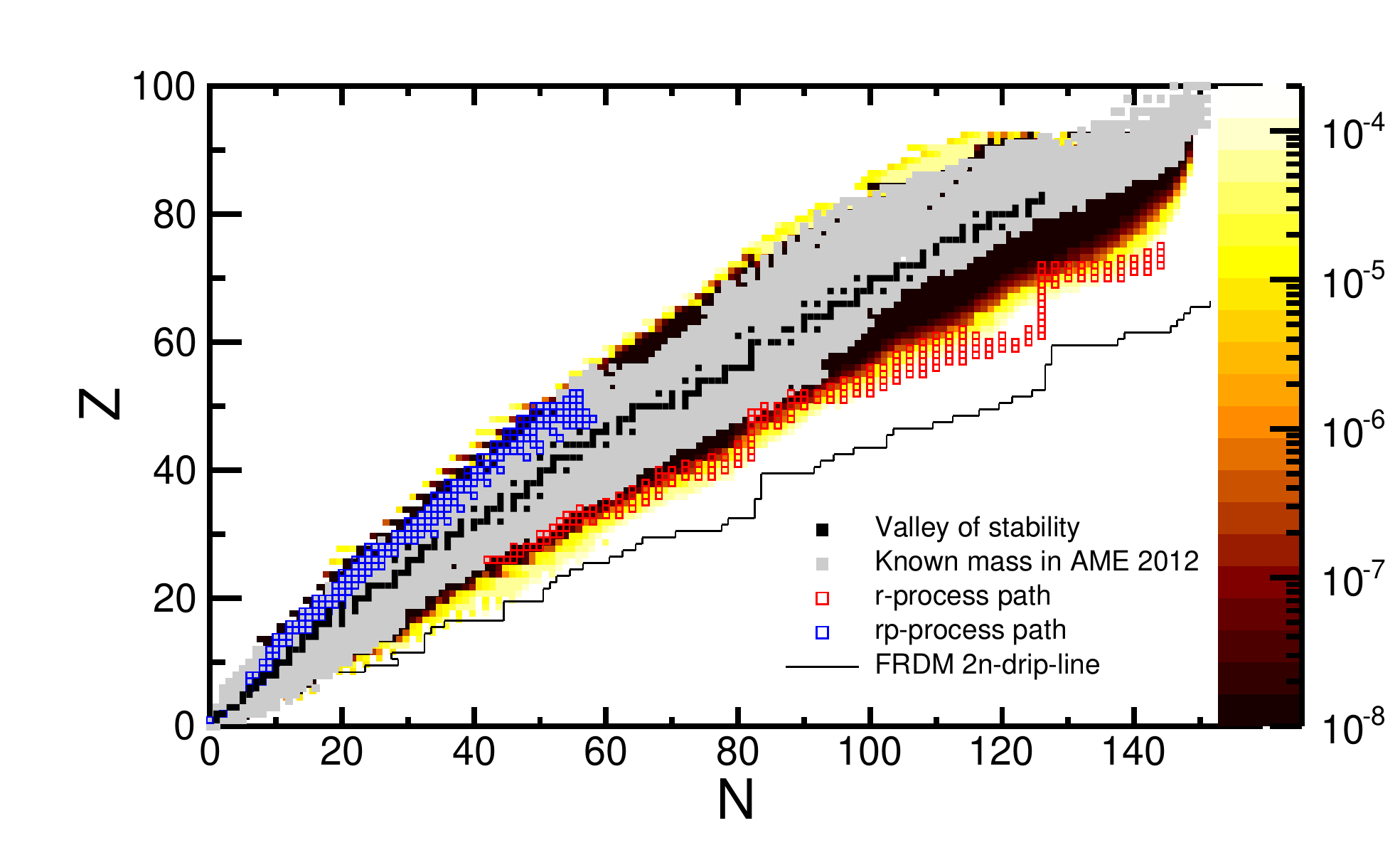}
\caption{Anticipated relative uncertainty $\delta m/m$ 
achievable by the combined use of PTMS and the TOF-$B\rho$ method at FRIB. The $rp$ and $r$-process paths are the same as those referred to in Figure~\ref{figure:exoticity}.
Note that life-time reductions due to
particle emission are neglected, and therefore estimates near the
proton drip-line are not reliable.}
\label{figure:chart}
\end{figure}

\subsection{Estimated precision for FRIB mass measurements}
\label{MassPrecision}

Figure~\ref{figure:chart} shows the predicted mass measurement
precision that can be achieved by the combined use of PTMS and
TOF-$B\rho$ at FRIB. Relative to the 1,098 neutron-rich masses reported in the 2012 Atomic Mass Evaluation~\cite{Audi12}, the predictions shown correspond to 965
higher-precision nuclear masses
and 1,172 new nuclear
masses on the neutron-rich side of stability: 693 from PTMS with a precision $\leq10^{-6}$ and 479 from
TOF-$B\rho$ with a precision $\leq10^{-4}$. These measurements would 
roughly double the known mass surface for neutron-rich nuclides, leading to advances in nuclear structure and
nuclear astrophysics. Masses will likely be obtained very near to the
neutron drip-line up to roughly iron and will
elucidate the evolution of the $N=82$ and (especially) $N=126$
shell-closures for decreasing proton numbers. The expansion of the
mass surface up to $A\approx100$ will tightly constrain the possible
strength of nuclear heating and cooling in the crusts of accreting
neutron stars~\cite{Meis18}. Whereas the expansion for the isotopes
above iron may deliver nuclear masses along the majority of the $r$-process
path, possibly distinguishing between hot and cold $r$-process
sites, for example with the neodymium masses~\cite{Mump16b}.

\section{Conclusions}

The TOF-$B\rho$ method has played and will continue to play a key role in mass spectrometry for exotic nuclides. Over the past three decades, such measurements have made significant contributions to our understanding of nuclear structure and nuclear astrophysics, particularly invovling neutron-rich nuclides. Though lower precision than other available methods, TOF-$B\rho$ measurements continue to map the frontiers of the nuclear mass surface.

\ack
I thank my many TOF-$B\rho$ mass measurement collaborators, especially Sebastian George, Alfredo Estrad\'{e}, Mike Famiano, Wolfi Mittig, Fernando Montes, Hendrik Schatz, and Dan Shapira. This work was supported in part by the U.S. Department of Energy Office of Science under Grants No. DE-FG02-88ER40387 and DESC0019042 and the U.S. National
Science Foundation through Grant No. PHY-1430152
(Joint Institute for Nuclear Astrophysics -- Center for
the Evolution of the Elements).

\section*{References}
\bibliographystyle{iopart-num}
\bibliography{References}

\end{document}